\documentclass[prl, superscriptaddress, twocolumn, showpacs, aps, floatfix, nofootinbib]{revtex4-1}

\usepackage{amsfonts}
\usepackage{amssymb}
\usepackage{graphicx}
\usepackage{color}
\usepackage{amsmath}
\usepackage[bookmarksnumbered, bookmarks, breaklinks, linktocpage]{hyperref}

\newcommand{\Tr}        {\mathrm{Tr}}

\newcommand{\ket}[1]    {| #1 \rangle}
\newcommand{\bk}[2]     {\langle #1 | #2 \rangle}
\newcommand{\kb}[2]     {| #1 \rangle \! \langle #2 |}
\newcommand{\cH}        {{\mathcal H}}
\newcommand{\cS}        {{\mathcal S}}
\newcommand{\cA}        {{\mathcal A}}
\newcommand{\cE}        {{\mathcal E}}

\newcommand{\eend}      {\hspace{\stretch{1}}\rule{1ex}{1ex}}

{%
 \definecolor{BLACK}{gray}{0}
 \definecolor{WHITE}{gray}{1}
 \definecolor{RED}{rgb}{1,0,0}
 \definecolor{GREEN}{rgb}{0,1,0}
 \definecolor{BLUE}{rgb}{0,0,1}
 \definecolor{CYAN}{cmyk}{1,0,0,0}
 \definecolor{MAGENTA}{cmyk}{0,1,0,0}
 \definecolor{YELLOW}{cmyk}{0,0,1,0}
 }

\newcommand\cF{{\mathcal F}}

\newcommand\dpcom[1]{}

\def\be{\begin{equation}}
\def\ee{\end{equation}}
\def\bea{\begin{eqnarray}}          
\def\eea{\end{eqnarray}}
\def\bi{\begin{itemize}}
\def\ei{\end{itemize}}

\usepackage{tikz}
\usetikzlibrary{arrows,shapes}

\newcommand\hocom[1]{}

\begin{document}

\tikzstyle{every picture}+=[remember picture]

\title{Quantum Darwinism, Decoherence, and the Randomness of Quantum Jumps}

\author{Wojciech H. Zurek}
\affiliation{Theory Division, LANL, MS B213, Los Alamos, NM  87545}


\begin{abstract} 
\noindent Tracing flows of information 
in our quantum Universe 
explains why we see the world as classical.
\end{abstract}

\maketitle

Quantum principle of superposition decrees every combination of quantum states a legal quantum state. This is at odds with our experience (Fig. 1). 

Decoherence 
selects preferred pointer states that survive interaction with the environment. 
They are localized and effectively classical. They persist while their superpositions decohere. Decoherence marks the border between quantum and classical, alleviating concern about flagrant and manifestations of quantumness in the macroscopic domain.

Here we consider emergence of `the classical' starting at a more fundamental pre-decoherence level, tracing the origin of preferred pointer states and deducing their probabilities from the core quantum postulates. We also explore role of the environment as a medium through which observers acquire information. This mode of information transfer leads to perception of objective classical reality.

\section{The Quantum Credo...}

Core quantum postulates are a strikingly simple and natural
and purely quantum 
part of a longer list of axioms found in textbooks \cite{Dirac}. They are behind `quantum weirdness', but we will see that they also help explain emergence of `the classical'.

Much of the weirdness follows from the {\it superposition principle} imp lied by postulate {\textbullet \bf 1}: {\it States of quantum systems correspond to normalized vectors in a (complex) Hilbert space,} 
$\ket s \in {\cal H}_{\cal S}$. Any superposition---any $\ket v = \alpha \ket r + \beta \ket s$, including `Schr\"odinger's cat'---is a legal state. 
Geometry of Hilbert spaces is euclidean, based on scalar products.
Pythagoras' theorem holds---when $\bk s r=0$, superposition satisfies ${\bk v v} =|\alpha|^2+|\beta|^2$. 

Core postulates include {\it unitarity} {\textbullet \bf 2}: {\it Evolutions are unitary}, $\ket {s_t } = {\bf U}_t \ket {s_0}$. Unitarity implies linearity (evolution yields superpositions of evolved ingredients, $\ket {v_t} = \alpha {\bf U}_t \ket r + \beta {\bf U}_t \ket s$) and preserves scalar products, $\bk {s_t} {r_t} = \bk {s_0} {r_0}$. Schr\"odinger equation can generate $ {\bf U}_t $.

{\it Composition} postulate {\textbullet \bf 0}: {\it State of a composite system is a vector in the tensor product of constituent Hilbert spaces,}  ${\cal H}_{\cal SA}={\cal H}_{\cal S}\otimes{\cal H}_{\cal A}$ is needed to deal with two or more systems (e.g., a measured system $\cS$ and a quantum apparatus $\cA$). Entanglement enters via postulate {\bf 0}.

Postulates {\bf 0-2} guide calculations involving ingredients such as Hamiltonians. But this is just `quantum math'.
To do physics one must relate math to experiments.

{\it Repeatability} {\textbullet \bf 3}: {\it Immediate repetition of a measurement yields the same outcome} starts this task. It is familiar -- measurements reveal classical states, so classical repeatability follows from objective existence of states. One cannot find out unknown quantum states, but repeatability lets one confirm presence of known states. 
Repeatability ties states' role as a summary of information \cite{Bohr} to their function as building blocks of reality.

Postulates {{\bf 0}-\bf 3} are our {\it quantum credo}. We will show they imply or at least motivate {\it measurement axioms}---the troubling remainder of the ``textbook list''.

\section{...and the measurement amendments}
The remaining textbook axioms deal with measurements, but, unlike postulate 3, are controversial. 

{\it Collapse axiom} {\bf 4} has two parts. {\textbullet \bf 4a}: {\it Observables are Hermitian} restricts outcome states {\it orthogonal} eigenstates. (Hermiticity of observables implies this orthogonality.) When the system is not in one of them, measurement cannot reveal its state. 
\textbullet{\bf 4b}: {\it One outcome is seen in each run} resets the state to an eigenstate specified by {\bf 4a} (allowing for repeatability, in accord with {\bf 3}).

{\begin{figure}[tb]
\vspace{-0.5cm}
\begin{tabular}{l}
\vspace{-0.2in} 
\includegraphics[width=3.5in]{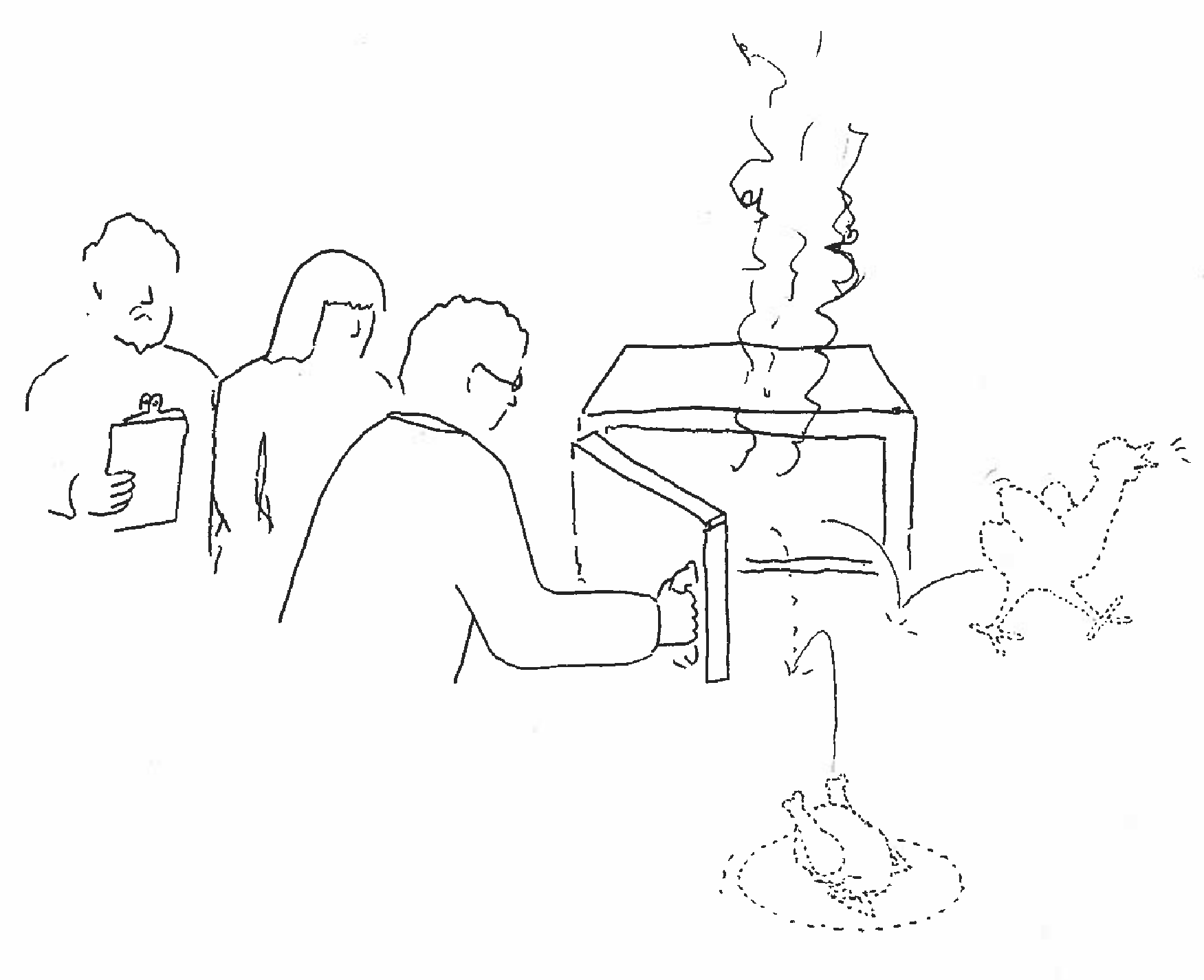}\\
\end{tabular} 
\caption{A great moment in the development of quantum microwave oven: Chicken \`a la Schr\"odinger. (Illustration by Ben Bromley.)
}
\label{chicken}
\end{figure}

Unknown classical states can be found out by many and remain the same. This is the evidence of their objective existence. By contrast, quantum systems are forced into outcome states specified by the eigenstate menu, axiom {\bf 4a}, which limits predictive value of quantum math. 

Finding out a quantum state without resetting it by direct measurement is ruled out.
However, initial state $\ket s$ determines probabilities of measurement outcomes via
{\it Born's rule}, \cite{Born} postulate {\textbullet \bf 5}: {\it Eigenstates $ \ket {o_k}$ of the observable $\hat {\bf O}=\sum_k \omega_k \kb {o_k} {o_k}$ are detected with the probability  $p_k=|\psi_k|^2$, square of the corresponding amplitude 
in the pre-measurement state $\ket s = \sum_k \psi_k \ket {o_k} $}. Born's rule is the key link between quantum math and physics.

Randomness stipulated by axioms {\bf 4,~5} clashes with the unitarity of {\bf 2}. 
Forefathers of quantum theory 
bypassed this conflict by insisting a part of the Universe (including measuring devices) must be classical. 
According to Niels Bohr \cite{Bohr}, selection of the menu of outcomes and randomness 
of `collapse' 
entered in this lawless border territory.

Revisiting foundations of quantum theory---the subject that preoccupied Bohr, Einstein, Dirac, von Neumann, and other forefathers is sometimes seen as almost disrespectful. Yet, there are implications of quantum theory relevant to its interpretation (such as Bell's inequalities, decoherence, and the no-cloning theorem) that were discovered long time after its inception. 

We will see that postulates {\bf 0-3} lead to axioms {\bf 4a} and {\bf 5}, as
the cross-border trade (information flows) must respect quantum constraints. Perception of objective reality arises when role of the environment 
as a communication channel that delivers our information is recognized.

\section{Repeatability and Quantum Jumps} 

Decoherence \cite{
Schlosshauer} 
leads to environment-induced superselection (einselection) of preferred states, accounting for quasiclassical states and the menu of measurement outcomes. 
However, reduced density matrices---key tool of decoherence---arise from averaging over the environment $\cE$, implicating Born's rule, axiom {\bf 5} (motivated by {\bf 4}). 

We will deduce
einselection and decoherence directly from postulates {\bf 0-3}, 
by considering information flows, bypassing controversial measurement axioms.

Consider a measurement-like interaction of a system $\cS$ with a quantum apparatus $\cA$. State of $\cA$ changes, but (to assure repeatability {\bf 3}) states of $\cS$ do not:
$$
\ket u \ket {A_0} { \stackrel {{\bf H}_{\cS\cA}} {\Longrightarrow} } \ket u \ket {A_u}, \ \ \ \ \  \ket v \ket {A_0}  { \stackrel {{\bf H}_{\cS\cA}} {\Longrightarrow} } \ket v \ket {A_v} \ . \eqno(1)
$$ 
As  $\ket u, \ket v$ are untouched
by the interaction ${{\bf H}_{\cS\cA}}$, 
another apparatus ${\cal A}'$ using analogous interaction ${\bf H}_{{\cS\cA}'}$ will get the same outcomes, in accord with the repeatability postulate {\bf 3}. 

The arrows in Eq. (1) represent a unitary evolution. Therefore, the ``before and after'' scalar products of state vectors in the composite Hilbert space ${\cH}_{\cS\cA}$ must equal:
$$
\bk u v \bk {A_0} {A_0} = \bk u v \bk {A_u} {A_v} \ . \eqno(2)
$$
This simple equation has profound consequences. We analyze them starting with a misstep: In an attempt  to simplify, we divide by $\bk u v$. We get $\bk {A_0} {A_0}=\bk {A_u} {A_v}$, or $\bk {A_u} {A_v}=1$. This implies $\ket {A_u} = \ket {A_v}$---apparatus can bear no imprint distinguishing $\ket u$ from $\ket v$! 

Have we just ruled out repeatable measurements---part of our credo---as incompatible with postulates {\bf 0-2}?
Not quite. Only when $\bk u v \neq 0$ can one `simplify' Eq. (2)!
Thus, {\it to be recorded repeatably states must be orthogonal}.  
This applies to any information flow (e.g. decoherence, where $\cE$ plays the role of $\cA$). Therefore, axiom {\bf 4a} follows from the core postulates (orthogonality of outcomes implies Hermiticity of observables \cite{Dirac}). Moreover, any $\bk {A_u} {A_v} \neq 1$ implies $\bk u v = 0$, so our assumptions are weaker than in {\bf 4a}, as the record quality does not matter.

Thus, one cannot find out an unknown state: repeatability and unitarity preclude it \cite{Zurek07}. By selecting interaction ${\bf H}_{\cS\cA}$ observers choose the measured observable $\hat {\mathbf \Lambda}$. Eigenstates of the observable that commutes with ${{\bf H}_{\cS\cA}}$;
$$ [\hat {\mathbf \Lambda}, {{\bf H}_{\cS\cA}}]=0  \eqno(3) $$
are preserved. Other states are disrupted. Unless observers agree beforehand what to measure, their measurements will disrupt one another and disagree, precluding discovery of an `objective reality' in a quantum Universe.

Pointer observables selected by decoherence satisfy similar condition
(see e.g. Physics Today {\bf 44}, 36, 1991).
Repeatability yields discrete outcomes setting stage for quantum jumps
({\it the} quantum trademark), 
and {\it perception} of collapse.
Discreteness arises from the conflict between linearity of quantum evolutions that is responsible for the no-cloning theorem, and nonlinearity of copying.

Born's rule, axiom {\bf 5}, was not used. Scalar products of 0 or 1 signified orthogonality and/or certainty. Their `in between' values were not needed. Thus, much of axiom {\bf 4} follows straight from the quantum core postulates {\bf 0-3}. 
The same is true---we now show---for decoherence.
Thus, we have deduced one of the measurement axioms from the quantum core.

Imperfect repeatability suffices \cite{Zurek07}. For instance, microstates of macroscopic memory (e.g., an apparatus pointer) can change, yet records they represent can be repeatably accessed and remain unperturbed as long as they correspond to orthogonal subspaces of the Hilbert space of the memory device, again implying discreteness (e.g., of the records of measurement outcomes).

We only needed a very rudimentary idea of information transfer: When $\ket {A_u} \neq \ket {A_v}$, {\it some} information is transferred to $\cA$. How much---it does not matter, at least for now (although we shall revisit this in detail later). This is fortunate, as we have not quantified information as yet. We shall do that only after we introduce probabilities.

For now, we have taken an important step towards probabilities: We have seen that, to be stable upon a re-measurement (as classical states should be), outcomes must be orthogonal, which means distinguishable---mutually exclusive. A sample space of mutually exclusive events is the foundation of probabilities \cite{Gnedenko}. Outcomes of repeatable measurement provide such a sample space.

\section{Decoherence as a Symmetry}

To study decoherence we change the recipient (before it was an apparatus, now the environment) and the initial state: We include superpositions of stable states. We shall see that, as before, states that survive repeated monitoring by the environment must be orthogonal, while phases of their coefficients become irrelevant for its state as they leak out from the system into its correlations with the environment. 

Decoherence is the loss of phase coherence between preferred states.
It occurs when $\cS$ starts in a superposition of pointer states singled out by the interaction, as in Eq. (1), but now $\cS$ is `measured' by $\cE$, its environment:
$$
(\alpha \ket \uparrow + \beta \ket \downarrow)\ket {\varepsilon_0} { \stackrel {{\bf H}_{\cS\cE}} \Longrightarrow } \alpha \ket \uparrow \ket {\varepsilon_\uparrow}  + \beta \ket \downarrow \ket {\varepsilon_\downarrow} = \ket {\psi_{\cS\cE}}. \eqno(4)
$$
Equation (2) implies that the untouched states are orthogonal, $\bk \uparrow  \downarrow = 0$. Their 
superposition 
turns into an entangled $\ket {\psi_{\cS\cE}}$: neither $\cS$ nor $\cE$ alone have a pure state.
This loss of purity signifies decoherence.
As we shall see, one can still assign a mixed state, encoding surviving information about $\cS$, to the system. 

Phases in superpositions matter: In a spin $\frac 1 2$--like $\cS$ $\ket \rightarrow = \frac {\ket \uparrow + \ket \downarrow} {\sqrt 2}$ is orthogonal to $\ket \leftarrow =\frac { \ket \uparrow - \ket \downarrow} {\sqrt 2}$.  Phase shift operator ${\bf u}^{\phi}_\cS=\kb \uparrow \uparrow +  e^{\i \phi}  \kb \downarrow \downarrow$ alters phase that distinguishes them: when $\phi=\pi$, it converts $\ket  \rightarrow $ to $\ket \leftarrow$. In experiments ${\bf u}_\cS^\phi$ would shift the interference pattern.

\hocom{In pure states phases matter; $\ket \rightarrow = \ket \uparrow + \ket \downarrow$ is orthogonal to $\ket \leftarrow = \ket \uparrow - \ket \downarrow$. One can adjust phases by acting, on $\cS$, with ${\bf u}_\cS^\phi=\kb \uparrow \uparrow +  e^{\i \phi}  \kb \downarrow \downarrow$. This phase shift operator converts $\ket  \rightarrow $ to $\ket \leftarrow$ when $\phi=\pi$. }

We assume perfect decoherence, $\bk {\varepsilon_\uparrow}{\varepsilon_\downarrow} = 0$: $\cE$ has a perfect record of pointer states.
What information survives decoherence, and what is lost?
We now show that when $\bk {\varepsilon_\uparrow}{\varepsilon_\downarrow} = 0$ 
phases of $\alpha$ and $\beta$ no longer matter for $\cS$---$\phi$ has no effect on {\it local} state of $\cS$, so measurements on $\cS$ cannot detect $\phi$---there is no interference pattern to shift.

Phase shift 
${\bf u}_\cS^\phi \otimes {\bf 1}_\cE$ acting on an entangled $\ket {\psi_{\cS\cE}}$
has no effect on its local state: It 
can be undone by ${\bf u}_{\cE}^{-\phi}=\kb {\varepsilon_\uparrow} {\varepsilon_\uparrow} + e^{-\i \phi} \kb {\varepsilon_\downarrow}{\varepsilon_\downarrow}$, a `countershift' acting on a distant $\cE$ decoupled from the system: 
$${\bf u}_{\cE}^{-\phi}({\bf u}_\cS^{\phi} \ket {\psi_{\cS\cE}})={\bf u}_{\cE}^{-\phi}(\alpha \ket \uparrow \ket {\varepsilon_\uparrow}  + e^{\i \phi}\beta \ket \downarrow \ket {\varepsilon_\downarrow})=\ket {\psi_{\cS\cE}} . \eqno(5) $$ 
As phases in $\ket {\psi_{\cS\cE}}$ can be changed in a faraway $\cE$
decoupled from, but entangled with $\cS$,
 they can no longer influence local state of $\cS$. (If they could, measuring $\cS$ would reveal this, enabling superluminal communication!)

We conclude that phases of the coefficients $\alpha, \beta$ are no longer relevant for the state of $\cS$: Phase coherence between the stable states of $\cS$ is lost. This happened as a result of the environment acquiring perfect records of the stable states---$\bk {E_u}{E_v} = 0$. 
Loss of phase coherence is decoherence. 

Superpositions of $\ket \uparrow, \ket \downarrow$ 
decohere as $\ket \uparrow, \ket \downarrow$ are recorded by $\cE$. 
This is not because phases become ``randomized'' by interactions with $\cE$, as is sometimes said. Rather, phase information becomes delocalized, so they lose significance for $\cS$ alone. They 
no longer belong to $\cS$ alone, so measurements on $\cS$ cannot distinguish states that started as superpositions with different phases for $\alpha, \beta$. 
Hence, information is also lost from $\cS$. We saw it here without reduced density matrices, the usual tool of decoherence.

Rigorous proof of 
the coherence loss \cite{Zurek03a} uses our credo and {\it facts (i)-(iii)};
{\it  (i) Locality: A unitary must act on a system to change its state.} State of $\cS$ that is not acted upon doesn't change even as other systems evolve (so phase shift ${\mathbf 1}_\cS \otimes (\kb {\varepsilon_\uparrow} {\varepsilon_\uparrow} + e^{-\i \phi} \kb {\varepsilon_\downarrow}{\varepsilon_\downarrow})$
that changed $\ket \rightarrow$ to $\ket \leftarrow$ in a pure $\cS$, 
does not affect $\cS$ when ${\cS\cE}$ are entangled, in $\ket {\psi_{\cS\cE}}$); {\it (ii) State of a system is all there is to predict measurement outcomes}; {\it (iii) A composite state determines states of subsystems} (so local state of $\cS$ is restored when the state of the whole $\cS\cE$ is restored). 

{\it Facts} help characterize local states of entangled systems without using density matrices. 
Thus, phase shift ${\bf u}_\cS^\phi \otimes {\mathbf 1}_\cE=(\kb \uparrow \uparrow +  e^{\i \phi}  \kb \downarrow \downarrow)  \otimes {\mathbf 1}_\cE$ 
acting on pure pre-decoherence states (Eq. (4)) matters:  
${\bf u}_\cS^\phi$ changes $\alpha \ket \uparrow$ $ + \beta \ket \downarrow$ into $\alpha \ket \uparrow+e^{\i \phi}  \beta \ket \downarrow$. 
However, the same ${\bf u}_\cS^\phi$ acting on $\cS$ in an entangled state $\ket {\psi_{\cS\cE}}$ does not matter for $\cS$ alone, as it can be undone by ${\mathbf 1}_\cS \otimes (\kb {\varepsilon_\uparrow} {\varepsilon_\uparrow} + e^{-\i \phi} \kb {\varepsilon_\downarrow}{\varepsilon_\downarrow})$, a countershift ${\bf u}_{\cE}^{-\phi}$ 
acting on a faraway, decoupled $\cE$. As the global $\ket {\psi_{\cS\cE}}$ is restored, by fact (iii) local state of $\cS$ is also restored even if $\cS$ is not acted upon (so by fact (i), it remains unchanged). As this restoration happens without any local evolution of $\cS$, 
the local state of decohered $\cS$ that obtains from $\ket {\psi_{\cS\cE}}$ 
could not have changed to begin with, and so it cannot depend on phases of $\alpha, \beta$.

This view decoherence appeals to symmetry; it invokes invariance of $\cS$---{\it en}tanglement-assisted in{\it variance} or {\it envariance} \cite{Zurek03a}---under phase shifts of the pointer state coefficients. As $\cS$ entangles, here with $\cE$, its local state becomes invariant under transformations that affected it beforehand. The only pure states invariant under such phase shifts 
$(\kb \uparrow \uparrow +  e^{\i \phi}  \kb \downarrow \downarrow ) 
\otimes {\mathbf 1}_\cE$ 
are the pointer states identified earlier, Eqs. (1,2), via their stability under information flows.

Preferred pointer states preserve correlations when monitored by $\cE$.
For instance, interaction of the measured $\cS$ with the apparatus, Eq. (1), leads to an entangled state that decoheres when $\cA$ interacts with $\cE$:
$$(\alpha \ket \uparrow \ket {A_\uparrow}  + \beta \ket \downarrow \ket {A_\downarrow} ) \ket {\varepsilon_0}
{ \stackrel {{\bf H}_{\cA\cE}} \Longrightarrow }
\alpha \ket \uparrow \ket {A_\uparrow} \ket {\varepsilon_\uparrow}  + \beta \ket \downarrow \ket {A_\downarrow} \ket {\varepsilon_\downarrow} = \ket {\Psi_{\cS\cA\cE}} \eqno(6)$$
Pointer states $\ket {A_\uparrow}, \ket {A_\downarrow}$ of $\cA$ survive decoherence by $\cE$. They retain correlations with the measured $\cS$ (or an observer, or other systems) in spite of $\cE$, independently of the value of $\bk {\varepsilon_\uparrow} {\varepsilon_\downarrow}$. Stability under decoherence is---in our quantum Universe---a prerequisite for classicality: classical states of macroscopic objects also have to survive monitoring by $\cE$ and, hence, retain correlations.

Decohered $\cS\cA$ is described by a {\it reduced density matrix}, $ \rho_{\cS\cA} = \Tr_\cE \kb {\Psi_{\cS\cA\cE}}{\Psi_{\cS\cA\cE}}$. When $\bk {\varepsilon_\uparrow} {\varepsilon_\downarrow} = 0$, it retains correlations of the pointer states of $\cA$ with the outcomes:
$$ \rho_{\cS\cA} 
= |\alpha|^2 \kb \uparrow \uparrow \kb {A_\uparrow} {A_\uparrow} + |\beta|^2 \kb \downarrow \downarrow \kb {A_\downarrow}{A_\downarrow} \eqno(7) $$
There is no `collapse'---all possible outcomes are present. 

Trace is a mathematical operation, but regarding e.g. the reduced density matrix $\rho_{\cS\cA}$ as a statistical mixture of its eigenstates with probabilities $|\alpha|^2,  |\beta|^2$ relies on 
Born's rule, axiom {\bf 5}. We abstained from using it so far to avoid circularity, but now we can derive it: Probabilities and reduced density matrices 
such as $\rho_{\cS\cA}$ 
will help quantify information and its flows.

Einselection selects pointer states. They are on the menu, but in the end a single outcome---one pointer state---is seen. 
We shall now derive its probability.

{\begin{figure}[tb]
\begin{tabular}{l}
\vspace{-0.15in} 
\includegraphics[width=3.5in]{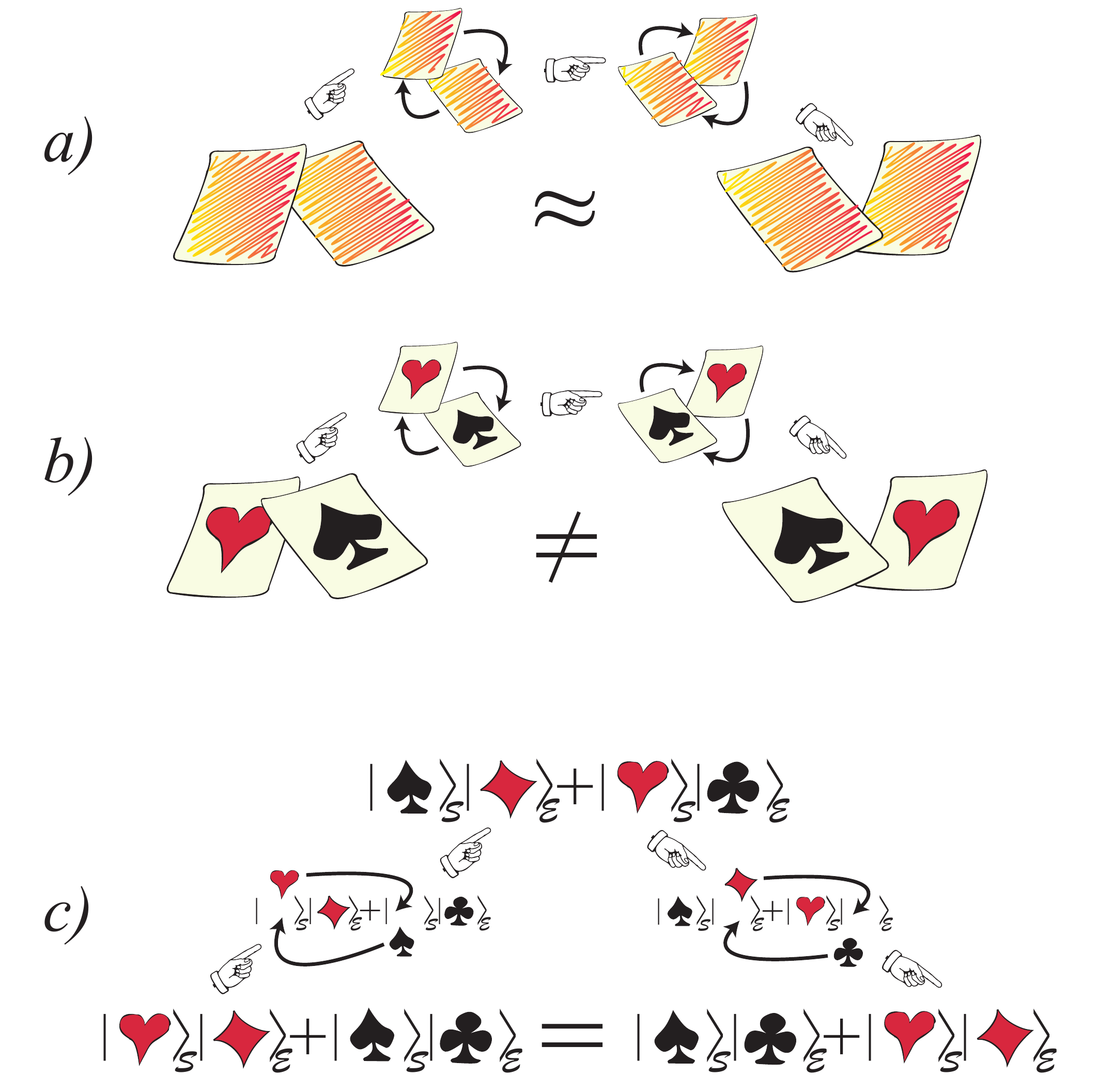}\\
\end{tabular}
\caption{Probability from entanglement. (a) Swapping (shuffling cards) illustrates subjective ignorance of a player: His indifference suggest equiprobability---symmetry
Laplace used to define probability. (b) Subjective ignorance is a shaky foundation: physical states change under swaps. (c) In a quantum case equiprobability follows from entanglement \cite{Zurek03a}: a swap in $\cS$ is undone by a counterswap in a distant $\cE$, proving that a swap in $\cS$ did not alter local state (including probabilities). Equiprobability follows from objective quantum symmetry, and leads to Born's rule. {\it Tensor structure} of pure entangled states of composite systems (that contrasts with {\it Cartesian product structure} of classical pure states, where every subsystem has its own pure state when the whole is pure) is essential. (Illustration by Fernando Cucchietti.)
}
\label{allcards}
\end{figure}

\section{Born's Rule from Entanglement}

Consider an observer who knows initial states of $\cS$ and $\cA$, interaction $\mathbf H_{\cS\cA}$ (hence, entangled state of $\cS\cA$), but not yet the outcome. 
What is its probability? 
We now use symmetry of entangled states to deduce probabilities.

Laplace developed probability theory starting with equiprobable alternatives  \cite{Gnedenko}.
According to his {\it principle of indifference}, when nothing favors any one outcome, symmetry implies they are equiprobable. Thus, probability of drawing blindly a spade is $\frac 1 4$ when nothing favors any suit
(e.g., for someone ignorant of the card order), 
because deck has 4 suits.
Indifference to swaps (Fig. 2) was regarded as invariance, evidence of symmetry. 
In the classical case such symmetry is due to {\it subjective} ignorance---it is in the mind of the beholder: 
The card on top either is or is not a spade. Classically, there is no {\it physical} basis for symmetry, and, hence, for equiprobability. 
Indeed, deducing probability from ignorance is circular: Probability is a way to quantify ignorance. 
The first step---proving equiprobability---is the hard one. 

In quantum physics one seeks probabilities of measurement outcomes starting from known states of $\cS$ and $\cA$. As pure entangled state of the whole is known, there is no ignorance in the usual sense. However, {\it en}tanglement-assisted in{\it variance} or {\it envariance} implies alternative outcomes with certifiably---objectively---equal probabilities.

Suppose $\cS$ starts as $\ket \rightarrow =\frac { \ket \uparrow + \ket \downarrow} {\sqrt 2}$, so interaction with $\cA$ yields $\frac { \ket \uparrow \ket {A_\uparrow}  + \ket \downarrow \ket {A_\downarrow}} {\sqrt 2}$, an {\it even} (equal coefficient) state 
(and we can skip normalization to save on notation).

Unitary {\it swap} $ \kb \uparrow \downarrow + \kb \downarrow \uparrow$ permutes states in $\cS$:

\vspace{-.3cm}

\[
\tikz[baseline]{
            \node[fill=gray!20,anchor=base] (t1)
            {$|\uparrow \rangle$};
        } 
| {A_\uparrow} \rangle 
+
\tikz[baseline]{
            \node[fill=gray!20,anchor=base] (t2)
            {$|\downarrow \rangle$};
        }
|{A_\downarrow} \rangle
\quad \longrightarrow \quad
| \downarrow\rangle |{A_\uparrow} \rangle + |\uparrow  \rangle | {A_\downarrow}\rangle . \ \eqnum ((8a)
\]
\begin{tikzpicture}[overlay]
        \path[->] (t1) edge [bend left=40] (t2);
        \path[->] (t2) edge [bend left=40] (t1);
\end{tikzpicture}

\vspace{-.3cm} 

\noindent 
After the swap $ \ket \downarrow $ is as probable as $\ket {A_\uparrow}$ was (and still is), and $\ket \uparrow$ as $\ket {A_\downarrow}$.  Probabilities in $\cA$ are unchanged (by fact (i)), so $p_\uparrow$ and $p_\downarrow$ must have been swapped. To prove equiprobability we now counterswap records in $\cA$:

\vspace{-.3cm}

\[
|\downarrow\rangle \tikz[baseline]{
            \node[fill=gray!20,anchor=base] (t3)
            {$| {A_\uparrow}\rangle$};
        } 
+
|\uparrow \rangle
\tikz[baseline]{
            \node[fill=gray!20,anchor=base] (t4)
            {$|{A_\downarrow} \rangle$};
        }
\quad \longrightarrow \quad
|\downarrow\rangle |{A_\downarrow} \rangle| + |\uparrow \rangle | {A_\uparrow}\rangle . \ \eqnum ((8b)
\]
\begin{tikzpicture}[overlay]
        \path[->] (t3) edge [bend left=40] (t4);
        \path[->] (t4) edge [bend left=40] (t3);
\end{tikzpicture}

\vspace{-.3cm}

\noindent  Swap in $\cA$ restores pre-swap state $ \ket \uparrow \ket {A_\uparrow}  + \ket \downarrow \ket {A_\downarrow}$ 
without touching $\cS$. 
Swap in $\cS$ followed by counterswap in $\cA$ restores the initial state, 
so (by fact (iii)) the local state of $\cS$ is also restored, and, by fact (ii), all predictions about $\cS$, {\it including probabilities} must be the same!

Probability of $\ket \uparrow$ and $ \ket \downarrow$, (as well as of $\ket {A_\uparrow} $ and $\ket {A_\downarrow}$) are exchanged yet unchanged. Therefore, 
for an even state 
they must be equal. Thus, in our two state case 
$p_\uparrow=p_\downarrow= \frac 1 2$. For $N$ envariantly equivalent alternatives, $p_k= \frac 1 N\ \forall k$ \cite{Zurek03a}.

Instead of subjective ignorance \`a la Laplace we used {\it objective quantum symmetries}: 
As in the uncertainty principle (where knowing $x$ precludes knowing $p$), indeterminacy of an outcome is a consequence of knowing something else---the whole entangled state. Objective indeterminacy of its parts ($\cS$ or $\cA$) and equiprobability of $\ket \uparrow$ and $\ket \downarrow$ follow. So, when an entangled state is even---amplitude of all pointer states are the same or they differ only by a phase (which is locally insignificant, as discussion of decoherence showed) their probabilities are equal! 

Using entanglement---quantum ingredient classical physics does not have---we proved something seemingly obvious: states with coefficients of equal magnitude have equal probability. (Readers who ``always thought so'' are forgiven for their impatience.) Yet,  this is the crux of the derivation: it establishes objective probabilities of a set of mutually exclusive alternatives, and leads to Born's rule in general. 
We established what Laplace, working in classical setting, could not get: Objective equiprobability based on physics---on core quantum postulates.

In an ``uneven'' $\alpha\ket \uparrow \ket {A_\uparrow}  + \beta \ket \downarrow \ket {A_\downarrow}$ swaps on $\cS$ and $\cA$ yield $\beta \ket \uparrow \ket {A_\uparrow}  + \alpha \ket \downarrow \ket {A_\downarrow}$, and not the pre-swap state, so $p_\uparrow$ and $p_\downarrow$ are not equal. However, this uneven 
case reduces to equiprobability via {\it finegraining}, so equiprobability in even states leads to Born's rule in general \cite{Zurek03a}: Let $\alpha \propto \sqrt \mu, ~ \beta \propto \sqrt \nu $, and $\mu, \nu$ be natural numbers. To finegrain, we change the basis; $\ket {A_\uparrow}=\sum_{k=1}^\mu \ket {a_{ k}}/\sqrt \mu$, and $ \ket {A_\downarrow}=\sum_{k=\mu+1}^{\mu+\nu} \ket {a_{ k}}/\sqrt \nu$, in the Hilbert space of $\cA$:
\vspace{-0.2cm}
$$
 \ket {\varphi_{\cS\cA} } \propto \sqrt \mu ~ \ket \uparrow \ket {A_\uparrow} + \sqrt \nu ~ \ket \downarrow \ket {A_\downarrow} $$
\vspace{-0.8cm}
 $$
= \sqrt \mu ~  \sum_{k=1}^\mu \ket \uparrow\ket {a_{ k}}/\sqrt \mu + \sqrt \nu ~ \sum_{k=\mu+1}^{\mu+\nu}\ket \downarrow \ket {a_k}/\sqrt \nu \ . \eqno(9a) 
$$ 
We simplify, and imagine an environment decohering $\cA$ in the new basis. That is, $\ket {a_k}$ correlate with $\ket {e_k}$
$$
\ket {\Phi_{\cS\cA\cE} } \propto\sum_{k=1}^\mu \ket {\uparrow {a_k}} \ket {e_k}+ \sum_{k=\mu+1}^{\mu+\nu}\ket {\downarrow {a_k} } \ket {e_k}  \eqno(9b)
$$
as if $\ket {a_k}$ were the preferred pointer states. Now swaps of $\ket {\uparrow {a_k}}$ with $\ket {\downarrow {a_k} } $ can be undone by counterswaps of $\ket {e_k}$.
Counts of the finegrained equiprobable ($p_k=\frac 1 {\mu + \nu}$) alternatives labelled with $\uparrow$ or $\downarrow$ lead to Born's rule:
$$ p_\uparrow = \frac \mu {\mu + \nu} =|\alpha|^2, \ \ \ \ p_\downarrow = \frac  \nu {\mu + \nu} = |\beta|^2 . \eqno(10)$$
Amplitudes `got squared' as a result of Pythagoras' theorem (euclidean nature of Hilbert spaces). Continuity settles the case of incommensurate $|\alpha|^2$ and $ |\beta|^2$. 

To prove equiprobability we used our credo and {\it facts} {\it (i)-(iii)}, invoked earlier to establish decoherence. Indeed, phase invariance can be used more directly: an even state $\ket \uparrow \ket {A_\uparrow}  + \ket \downarrow \ket {A_\downarrow}= \ket \rightarrow \ket {A_\rightarrow } + \ket \leftarrow \ket {A_\leftarrow} $ can be swapped by a phase shift in a complementary basis, as $\kb \rightarrow \rightarrow - \kb \leftarrow \leftarrow =  \kb \uparrow \downarrow + \kb \downarrow \uparrow$. There is an appropriate countershift (counterswap) in $\cA$, so dealing with phases suffices to establish equiprobability.

Envariance provides insight into quantum probability missing in the measure-theoretic theorem of Gleason \cite{Gleason}. The strategy---counting of equiprobable alternatives rather than past events---differs from more common (but unsuccessful in quantum setting \cite{Weinberg}) relative frequency approach. Envariance can be used to derive frequencies, and even shows that amplitudes must be proportional to their square roots, `inverting' Born's rule  \cite{Zurek03a}.

\section{Information Interlude}

Our focus so far was on quantum postulates. Textbooks start with lists that include uncontroversial core (quantum math plus repeatability) along with controversial measurement axioms. 
Measurement happens on the quantum-classical border where---one is told---laws of neither realm apply. Study of quantum measurements is often discouraged by such `manuals'. 
Prohibition on the study of the consistency of the postulates was a part of the instruction manual. 

Decoherence ignores that prohibition. It 
builds on von Neumann's \cite{vonN} analysis of measurements (that predated most textbooks), but begins to recognize role of the environment. However, its usual practice relied on Born's rule, axiom {\bf 5}, to justify physical significance of reduced density matrices. We now have a fundamental justification of this practice. 
Moreover, we shall see that environment does more than decohere: It acts as a communication channel through which we acquire our information.

Advances discussed above started with decoherence, but go beyond it. We shored up quantum foundations by re-deriving Born's rule and einselection (axiom {\bf 4a}).
The four cornerstones we built on---core quantum postulates {\bf 0-3}---account for much of what textbooks put in with controversial measurement axioms. 

We shall now show how objective classical reality arises in our quantum Universe. We already have a fundamental justification that allows us to use density matrices as they were always used---as statistical mixtures of their eigenstates---to calculate entropy and information needed in the study of quantum Darwinism. Below, we shall also reconsider the role of the decohering environment.

Pointer states preserve correlations, as evident in $ \rho_{\cS\cA}$, Eq. (7). This one-to-one correspondence of states of $\cS$ and $\cA$ does not rely on Born's rule. However, to quantify information $\cA$ has about $\cS$ we need probabilities $p_\uparrow=p_{A_\uparrow}=|\alpha|^2$,  
$ p_\downarrow=p_{A_\downarrow}=|\beta|^2$. 

Entropies of $\cS$, $\cA$, and the composite $\cS\cA$ are given by the von Neumann entropy \cite{vonN}, $H(\rho)= - \Tr \rho \lg \rho$. It generalizes classical $-\sum p_k\lg p_k$ and often, after decoherence, coincides with it. They are all equal for $ \rho_{\cS\cA}$ of Eq. (7);
$$H_\cS=H_\cA=H_{\cS\cA}=-(|\alpha|^2\lg |\alpha|^2 + |\beta|^2\lg |\beta|^2) . \eqno(11)$$
This means $\cS$ and $\cA$ know each other's preferred states perfectly. Two copies of a book share such information: each copy reveals content of both. 

{\it Mutual information}:
$$I(\cS : \cA)=H_\cS+H_\cA-H_{\cS\cA} \eqno(12)$$
quantifies how much two systems know about each other \cite{Cover}. For uncorrelated  $\rho_{\cS\cA}=\rho_{\cS}\rho_{\cA}$, $H_{\cS\cA}=H_\cS+H_\cA$, so $I(\cS : \cA)=0$.
For $\rho_{\cS\cA}$ of Eq. (7), $I(\cS : \cA)=H_\cS=H_\cA$
saturates classical limit, $I(\cS : \cA) \le \min(H_\cS,H_\cA)$ (information shared cannot exceed  smaller book's content).

Quantum correlations can be stronger; entanglement correlates every basis, $ \frac {\ket {\uparrow} \ket {  \uparrow }+ \ket {\downarrow} \ket { \downarrow} } {\sqrt 2}= \frac{\ket {\rightarrow } \ket {\rightarrow} + \ket {\leftarrow} \ket { \leftarrow}} {\sqrt 2}$. Decoherence that selects pointer states $\ket \uparrow, \ket \downarrow$ yields 
a mixture $\frac { \kb \uparrow \uparrow \kb \uparrow \uparrow + \kb \downarrow \downarrow \kb \downarrow \downarrow} 2$: Pointer states remain correlated, but $\ket \rightarrow, \ket \leftarrow$ that revealed pure states of the partner before now provide no such information. 
Mutual information reflects this: For pure entangled $\ket {\psi_{\cS\cE}}$, Eq. (4), $H_{\cS\cA}=0$, 
while $H_\cS=H_\cA=-|\alpha|^2\lg |\alpha|^2 - |\beta|^2\lg |\beta|^2$, so $I(\cS : \cA)=2 H_\cS$. Decohered $\rho_{\cS\cA}$, Eq. (7), has mutual information at the classical limit, $I(\cS : \cA)=H_\cS=H_\cA=H_{\cS\cA}$. 


Mutual information is the key tool of quantum Darwinism. When $I(\cS : \cA)=H_\cS$, an apparatus can fully reveal the state of $\cS$. In quantum Darwinism a fragment $\cF$ of the environment will play the role of the apparatus. Its correlation with the system will be often effectively classical, as the rest of the environment (denoted by $\cE\backslash\cF$) assures decoherence.

\section{Quantum Darwinism}

{\begin{figure}[tb]
\begin{tabular}{l}
\includegraphics[width=3.5in]{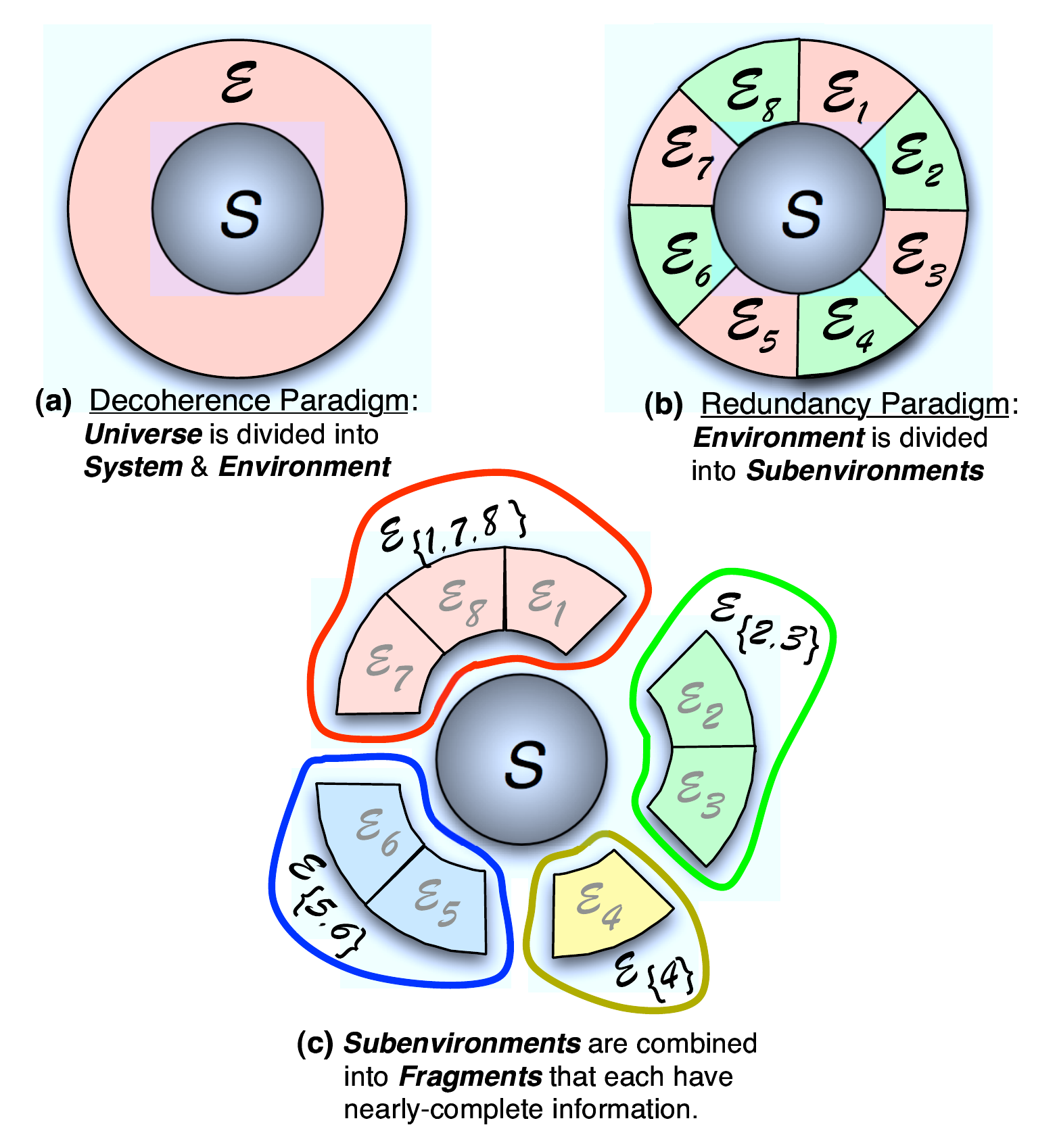}\\
\end{tabular}
\caption{Quantum Darwinism \cite{QD} recognizes that environments consist of many subsystems, and that observers acquire information about system of interest $\cS$ by intercepting copies of its pointer states deposited in $\cE$ as a result of decoherence.
}
\label{EnvSubdivision}
\end{figure}

Already einselection hints at survival of the fittest, as environments select pointer states that survive and aspire to classicality. Quantum Darwinism \cite{QD} goes further: In course of decoherence selected information proliferates into copies of pointer states of $\cS$ or $\cA$ imprinted on $\cE$. 

We monitor our world indirectly,
by eavesdropping on the environment. For instance, you are now intercepting a tiny 
fraction of photons scattered from this page. Anyone intercepting other fractions will see the same images. 

For many subsystems, $\cE=\bigotimes_k \cE^{(k)}$, so the initial state $(\alpha \ket \uparrow 
+ \beta \ket \downarrow) 
\ket { {\varepsilon^{(1)}_0} {\varepsilon^{(2)}_0} {\varepsilon^{(3)}_0}...}$ evolves (as in Eq. (4)), into: 
$$ \ket {\Upsilon_{\cS\cE}} = \alpha \ket \uparrow \ket { {\varepsilon^{(1)}_\uparrow} {\varepsilon^{(2)}_\uparrow}{\varepsilon^{(3)}_\uparrow}... } + \beta \ket \downarrow \ket { {\varepsilon^{(1)}_\downarrow} {\varepsilon^{(2)}_\downarrow}{\varepsilon^{(3)}_\downarrow}...} \eqno(13)$$
Linearity assures both branches are still there: collapse to a single outcome will not happen.
However, large $\cE$ can disseminate information about the system. The state $\ket {\Upsilon_{\cS\cE}}$ represents many records inscribed in its fragments, collections of subsystems of $\cE$ (see Fig. 3). 
This means that the state of $\cS$ can be found out by many, independently, and indirectly---hence, without disturbing $\cS$. 

Environment fragment $\cF$ can act as an apparatus with a (possibly incomplete) record of $\cS$. When $\cE \backslash \cF$, the rest of the $\cE$, is traced out, $\cS\cF$ decoheres, and the reduced density matrix describing joint state of $\cS$ and $\cF$ is:
$$ \rho_{\cS\cF} = \Tr_{\cE \backslash \cF} \kb {\Psi_{\cS\cE}}{\Psi_{\cS\cE}}= |\alpha|^2 \kb \uparrow \uparrow \kb {F_\uparrow} {F_\uparrow} + |\beta|^2 \kb \downarrow \downarrow \kb {F_\downarrow}{F_\downarrow} \eqno(14) $$
in close analogy with Eq. (7). When $\bk {F_\uparrow} {F_\downarrow} =0$, $\cF$ contains a perfect record of the preferred states of the system. In principle, each subsystem of $\cE$ may be enough to reveal its state, but this is unlikely. Typically, one must collect many subsystems of $\cE$ into $\cF$ to find out about $\cS$, e.g. to distinguish $\uparrow$ from $\downarrow$.

The number of copies of the data about pointer states in $\cE$ determines how many times the same information can be independently extracted---it is a measure of objectivity. 
The key question of quantum Darwinism is then {\it How many subsystems of $\cE$---what fraction of $\cE$---does one need to find out about $\cS$?}. The answer is provided by the mutual information 
$$I(\cS : \cF_f)=H_\cS + H_{\cF_f} - H_{\cS \cF_f} \eqno(15)$$ 
that is the information about $\cS$ available from $\cF_f$, a 
fraction $f= \frac { \sharp \cF } { \sharp \cE }$ of $\cE$ (where $\sharp \cF$ and $ \sharp \cE$ are the numbers of subsystems). 

{\begin{figure}[tb]
\begin{tabular}{l}
\includegraphics[width=3.5in]{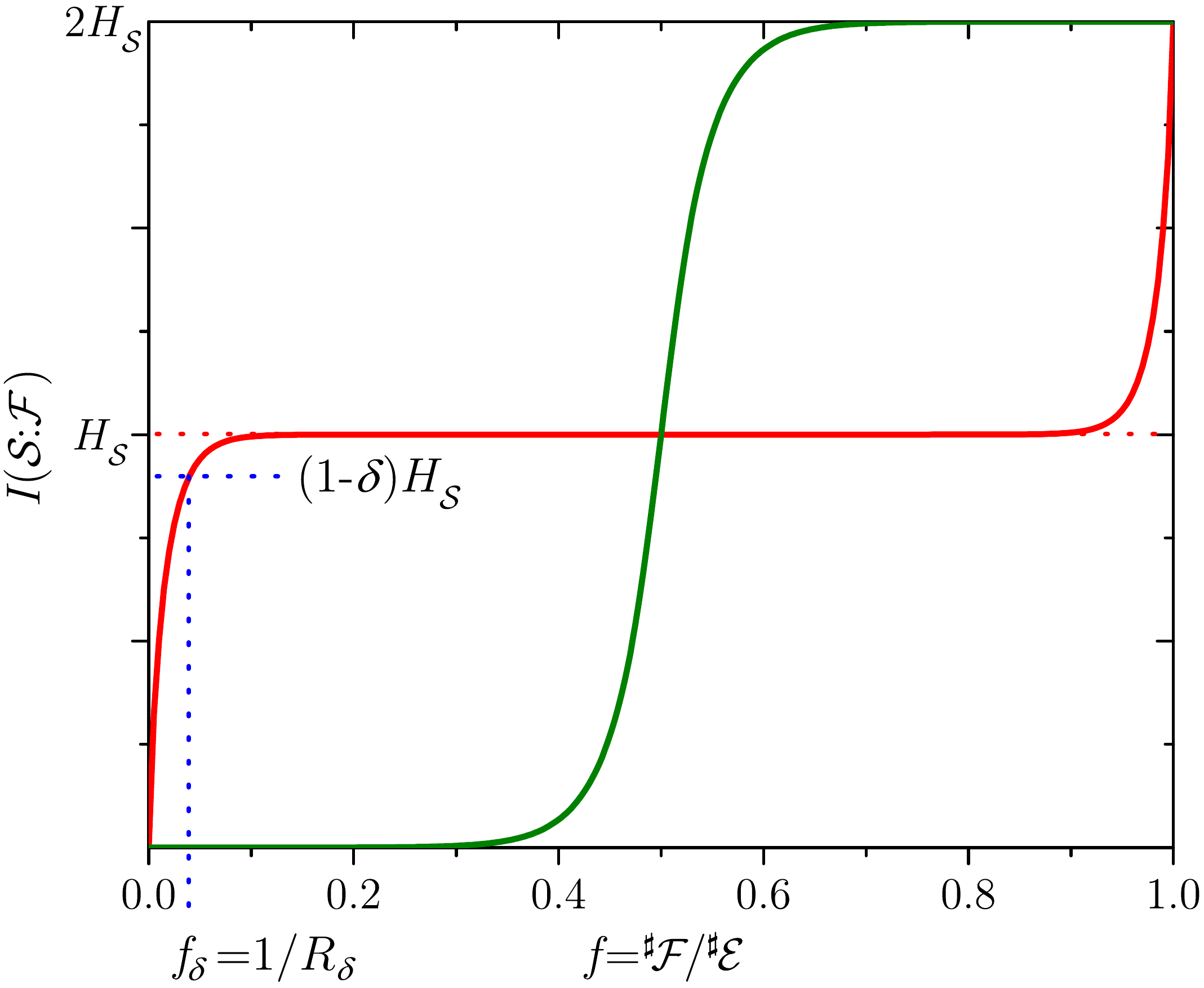}\\
\end{tabular}
\caption{Information about the system contained in a fraction $f$ of the environment. Red plot shows a typical $I(\cS : \cF_f)$ established by decoherence. Rapid rise means that nearly all classically accessible information is revealed by a small fraction of $\cE$. It is followed by a plateau: additional fragments only confirm what is already known. Redundancy ${\cal R_\delta} = 1 / f_\delta$ is the number of such independent fractions. Green plot shows $I(\cS : \cF_f)$ for a random state in the composite system ${\cS\cE}$.
}
\label{RedPIP}
\end{figure}

In case of perfect correlation a single subsystem of $\cE$ would suffice, as $I(\cS : \cF_f)$ jumps to $H_\cS$ at $f=\frac 1 {\sharp \cE}$. The data in additional subsystems of $\cE$ are then redundant---they confirm what observers already know.
Usually, however, larger fragment of $\cE$ is needed to find out about $\cS$. Red plot in Fig. 4 illustrates this: $I(\cS : \cF_f)$ still reaches $H_\cS$, but only gradually. The length of this plateau can be measured in units of $f_\delta$, the initial rising portion of $I(\cS : \cF_f)$. It is defined with the help of the {\it information deficit} $\delta$ observers tolerate:
$$ I(\cS : \cF_{f_\delta}) \ge (1- \delta) H_\cS \eqno(16) $$
Redundancy is the number of such records of $\cS$ in $\cE$:
$$ {\cal R_\delta} = 1 / f_\delta \eqno(17)$$
$ {\cal R_\delta}$ sets the upper limit on how many observers can find out the state of $\cS$ from $\cE$ independently and indirectly. In models \cite{QD} (e.g., decoherence due to photon scattering \cite{JoosZeh}) $\cal R_\delta$ can be huge \cite{Riedel}, and it depends on $\delta$ only weakly (logarithmically). 

This is `quantum spam': $\cal R_\delta$ imprints of pointer states are broadcast through the environment. Many observers can access this information independently and indirectly, assuring objectivity of pointer states of $\cS$. Repeatability is key: States must survive copying to produce many imprints. As we saw earlier, Eqs. (1)-(3), this is possible providing they are distinguishable---i.e., orthogonal \cite{Zurek07}.  

\section{Environment as a witness}

Not all environments are good in this role of a witness. Photons excel: They do not interact with the air or with each other, faithfully passing on information. Small fraction of photon environment usually reveals all we need to know. Scattering of sunlight builds up redundancy with time: a 1$\mu$ dielectric sphere in a superposition of similar size increases ${\cal R}_{\delta=0.1}$ by $ \sim 10^8$ every microsecond \cite{Riedel}. 

Air is also good in decohering, but its molecules interact, scrambling acquired data. Both air and photons scatter of the objects of interest, so both acquire information about position, and favor localized pointer states. 

Quantum Darwinism also shows why it is hard to undo decoherence. Plots of mutual information $I(\cS : \cF_f)$ for initially pure $\cS$ and $\cE$ are antisymmetric (see Fig. 4) around $f= \frac 1 2$ and $H_\cS$ \cite{QD}. Hence, a counterpoint of the initial quick rise at $f \le f_\delta$ is a quick rise at $f \ge 1 - f_\delta$, as last few subsystems of $\cE$ are included in the fragment $\cF$ that by now contains nearly all $\cE$. 
This is because an initially pure $\cS \cE$ remains pure under unitary evolution, so $H_{\cS \cE}=0$, and $I(\cS : \cF_f)|_{f=1}$ must reach $2 H_{\cS}$. Thus, a measurement on {\it all} of $\cS \cE$ could confirm its purity in spite of decoherence caused by $\cE \backslash \cF$ for all $f \le 1- f_\delta$. 

However, to verify this one has to intercept and measure all of $\cS\cE$ in a way that reveals pure state $\ket {\Upsilon_{\cS\cE}}$, Eq. (13). Other measurements destroy phase 
(i.e., quantum) 
information. So, undoing decoherence is in principle possible, but required resources and foresight preclude it.

In quantum Darwinism decoherence acts as an amplifier,
leading to branch structure of $\Upsilon_{\cS\cE}$. This state differs from typical states in the Hilbert space of $\cS\cE$: Random states have $I(\cS : \cF_f)$ given by the green plot in Fig. 4. There is now no plateau, and no redundancy. 
Antisymmetry still holds, so $I(\cS : \cF_f)$ ``jumps'' at $f= \frac 1 2$ to $2 H_{\cS}$. 

Environments that decohere $\cS$, but scramble information about it because of interactions between their subsystems (e.g., air) eventually approach such random states. Quantum Darwinism is possible only when information about $\cS$ is preserved in fragments of $\cE$, so that it can be recovered by local observers. However, there is no need for perfection: Partially mixed environments or imperfect measurements correspond to noisy communication channels: their capacity is depleted, but we can still get the message \cite{Zwolak}.

\section{Quantum Theory of Information and Classical Existence}

Our credo led to Hermiticity of observables and Born's rule, textbook axioms {\bf 4a} and {\bf 5}. Postulates {\bf 0-3} served as cornerstones to rebuild, simplify, and strengthen quantum foundations, and to explore quantum origins of information and perception of objective existence. 

Newtonian Universe had a strict separation of the `ontic' and `epistemic': states existed independently of information `about them'. 
Measurement `collapsed' observer's ignorance, leaving the state of the Universe untouched: information was unphysical, as if observer was outside of the Universe. There was no way to turn information into action---initial conditions preordained future evolution.

Quantum theory abolished this separation. Some miss information-independent objective existence: its absence ignites interpretational debates. John Bell \cite{Bell} was hoping for a theory of `{\it beables} that are {\it described in classical term, because they are there}'. We have seen instead how fragile quantum states can account for what we regard as ``objective existence''. 
Information is reflected in the state of the Universe.
Observers can choose what to measure. State of the measured system is modified along with observer's memory. Objective reality emerges when minute change inflicted on the quantum state by 
information gain---e.g., by a few photons---reveal 
a branch (that can be approximated by classical Cartesian product) sprouting from a pointer state. Additional measurements confirm what is already known, leaving an impression that we just find out what was there beforehand.

Repeatability leads to branch-like states, Eq. (13), suggesting Everettian `relative states' \cite{Everett}. There is no need to attribute reality to all the branches. Quantum states are part information. As we have seen, objective reality is an emergent property. Unobserved branches can be regarded as events potentially consistent with the initially available information that did not happen. Information we gather can be used to advantage---it can lead to actions conditioned on measurement outcomes \cite{Zurek07}.

John Wheeler, Charles Bennett, and others have considered the relation between 
information and existence \cite{Wheeler, Bennett}. Quantum Darwinism adheres to the quantum credo and adds to that discussion by recognizing that a decohering environment can be a communication channel. But since observers intercept only fractions of the environment, information about systems is only accessible when it is redundantly imprinted. Put another way, an observer can get information only about pointer states that remain intact despite monitoring by the environment: Using the it as a communication channel
comes at the price of censorship. Fractions of the environment reveal
branches one at a time and suggest quantum jumps.

Quantum Darwinism explains why we see only one branch. One can dismiss other branches, e.g. with an appeal to Everett \cite{Everett}. 
So we can account for a {\it perception} of collapse. 
Thus, while unitarity precludes fundamental collapse, local observables that reveal branches do not commute with the global observable whose eigenstates are coherent superpositions of all the branches, Eq. (13). Therefore, local observers have no way to probe (hence, cannot perceive) the global state vector. 

The basic tenets of decoherence have been confirmed
by experiment \cite{HWZ}. It may also be possible
to test quantum Darwinism; envariance is already
being tested \cite{Ver}.

Our proofs of Hermiticity, {\bf 4a} and of Born's rule, {\bf 5} are straightforward. They fit into the picture based on decoherence process that amplifies and disseminates information about selected (pointer) observables throughout the environment. Quantum Darwinism shows why only such redundantly recorded pointer states are accessible to observers---it can account for perception of `quantum jumps'. However, full account of collapse involves `consciousness', and may have go beyond just mathematics or physics. Good questions are valuable. It may yet turn out that residual worries about collapse lead to a good question. Of course, one only knows a question was good after it has led to an interesting answer. 
It seems unlikely but not impossible that the residual worries about quantum postulates can precipitate a similar breakthrough. 

 I thank Charles Bennett, Robin Blume-Kohout, Jim Hartle, Raymond Laflamme, Juan Pablo Paz, Hai-Tao Quan, Jess Riedel, Wolfgang Schleich, Max Tegmark, and Michael Zwolak for enjoyable and helpful discussions. Support by DoE (via LDRD at Los Alamos) and, in part, by FQXi is gratefully acknowledged.

\end{document}